\documentclass[article,english,twocolumn]{revtex4-1} 
\usepackage[T1]{fontenc} \usepackage[latin1]{inputenc}
 \usepackage{babel} \usepackage{txfonts}
 \usepackage{epsfig,epsf,psfrag} 
\usepackage{graphicx} 
\usepackage{subfigure}
\usepackage{pslatex}
\usepackage{epic,eepic} 
\usepackage{color,pstcol}
\usepackage{pstricks} 
\usepackage{fancyhdr}  \parindent0em  
\def\ua{\uparrow}
\def\da{\downarrow}

 \vspace{-10.0cm} 
\begin{document} \title{Adiabatically twisting  a magnetic molecule to generate pure spin currents in graphene} 
  \author{Firoz Islam}
 \author{Colin Benjamin} \email{cbiop@yahoo.com}\affiliation{ National institute of Science education \& Research, Bhubaneswar 751005, India }
\begin{abstract}
  The spin orbit effect in graphene is too muted to have any observable significance with respect to
  its application in spintronics. However,  graphene technology is too valuable to be rendered impotent to spin transport.
  In this communication we look at the effect of adiabatically twisting a single molecule magnet embedded in a graphene monolayer.
  Surprisingly, we see that pure spin currents (zero charge current) can be generated from the system via quantum pumping.
  In addition we also see spin selective current can also be pumped from the system. The pure spin current seen is quite resilient 
  to temperature while disorder has a limited effect. Further the direction of these spin pumped currents can be easily and exclusively controlled by the magnetization of the single molecule magnet with disorder having no effect on the magnetization control of the pumped spin currents. 
 \end{abstract}

\maketitle
\section{Introduction}
Graphene has been a revolutionary material of this the 21st century. Its remarkable that since its discovery 
around a decade ago it has captured the interest of the scientific community in ways that even High Tc superconductors 
in its hey days couldn't. A cursory look at graphenes properties would provide reasons for this. Graphene is light with
huge tensile strength and crucially in contrast to most materials, electronic energy is linearly proportional to its wave vector 
and not its square and it shows quite remarkable quantum phenomena like Klein tunneling\cite{novoselov} and room temperature
quantum Hall effect\cite{geim}, etc. Notwithstanding the many advantages of using graphene to transport electrons faster and with
less dissipation it has been observed that spin orbit effect in graphene is at best negligible thus making them of no practical
use for applications in spintronics. Then how do we exploit the manifest advantages of graphene and marry them to spintronics.
In this communication we address this issue. We intend to embed a monolayer of Graphene with a single molecule magnet.
We then adiabatically modulate two independent parameters of the resulting system to pump a pure spin current. 
The two parameters we modulate are one- the Magnetization strength of the magnetic molecule which has been shown
to be controlled by twisting it\cite{inglis}. Secondly, we add a delta function like point interaction, which can either be an             
 isolated adatom in the Graphene monolayer or an extended line defect or even a very thin potential barrier with the strength of the potential barrier being controlled by a gate voltage. Using these two parameters we intend to pump a pure spin current. 

\begin{figure}[h]
\begin{center}
\includegraphics[width=.5\textwidth,height=75mm]{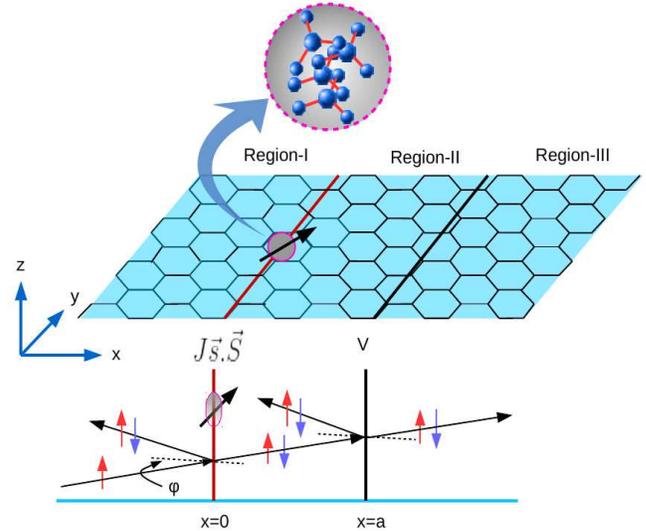}
\caption{A single molecule magnet at $x=0$ is adiabatically twisted along with an electrostatic delta potential at $x=a$ to 
generate pure spin currents in graphene. }
\label{Fig1}
\end{center}
\end{figure}

 Adiabatic quantum pumping is a means of transferring
charge and/or spin carriers without applying any external
perturbation (like voltage bias, etc.) by the cyclic variation
of two independent system parameters. The theory of
adiabatic quantum pumping was developed in 1998\cite{brouwer} based
on the fact that adiabatic modulations to a 2DEG can lead to
changes in local density of states and thus current transport.
In 1999, an adiabatic quantum electron pump was reported
in an open quantum dot where the pumping signal was produced
in response to the cyclic deformation of the confining
potential\cite{switkes}. The variation of the dot's shape changes electronic
quasi energy states of the dot thus `pumping' electrons
from one reservoir to another.

The AC voltages applied to the quantum dot in order to change the shape result in a DC
current when the reservoirs are in equilibrium. This non-zero current is only produced if there are at least two
time varying parameters in the system as a single parameter quantum pump does not transfer any charge.
Later on the study of adiabatic pumping phenomenon has been extended to adiabatic spin pumping both in 
experiment as well as theory\cite{benjamin,watson}. In the experiment on quantum spin pumping one generates a
pure spin current via a quantum dot by applying an in plane magnetic field which is adiabatically modulated to 
facilitate a net transfer of spin.  Many spintronic devices, such as the spin valve and magnetic tunneling 
junctions\cite{moodera}, are associated with the flow of spin polarized charge currents. Spin polarized currents
coexist with charge currents and are generated when an imbalance between spin up and spin down carriers is created,
for example, by using magnetic materials or applying a strong magnetic field or by exploiting spin-orbit coupling in
semiconductors\cite{moodera}. The spin transport aided by bias voltage have been studied earlier\cite{nature,prb_mhd,prb_zomer}.

More recently, there has been an increasing interest in the generation of pure spin 
current by quantum pumping without an accompanying charge current\cite{benjamin}.  In monolayer graphene,
a group has attempted to address this issue by proposing a device, consisting of ferromagnetic strips and 
gate electrodes\cite{lin}. However, ferromagnetic strips in graphene
are unwieldy and lead to further complication as spin polarization is lost at the feromagnet-graphene interface. In this work
we look at only a piece of monolayer graphene without any
ferromagnets but with two impurities one magnetic(SMM) the
other non-magnetic. Independently modulating these two we
show generation of pure spin currents. The generation of a pure spin-current is only possible 
if all spin-up electrons flow in one direction and equal amount of spin-down electrons flow in the opposite direction.
In this case the net charge current $I_{charge}=I_{\ua}+I_{\da}$ vanishes while a finite spin current $I_{spin}=I_{\ua}-I_{\da}$ exists,
because $I_{\ua}=-I_{\da}$, where $I_{\ua}$ or $I_{\da}$ are the electron current with spin-up or spin-down. 
Further we also see generation of spin selective currents namely either only $I_{\ua}$ or $I_{\da}$ and this implies the total
magnitude of charge current is identical to that of the spin current.

In this work we show how pure spin currents and spin selective currents can be generated via twisting a single molecule magnet. Single molecule magnets are of great importance for molecular spintronics. Such molecules can now be synthesized, 
for example the molecule $Mn_{12}$ac, which has a ground state with a large spin (S=10), and a very slow relaxation of magnetization at low temperature\cite{sessoli}. SMM's do not just have integer
spin values a number of SMM's like Mn4O3 complex\cite{wernsdorfer} have
large half integer spin S = 9/2. However, in a literal twist to
this tale (pun intended), a few years back scientists revealed
that these big spin molecules can be twisted and thus their
magnetization can be easily altered
\cite{brechin}. This fact we utilize in our adiabatic 
pumping mechanism.  The mechanism of twisting is not difficult to implement. SMM's have a metallic core (for example in $Mn_{12}$ complexes 
the metallic core consists of the $Mn-O-N-Mn$ moiety) and a deliberate targeted structural distortion of it leads to change in the magnetization
of the SMM so much so that a initially ferromagnetic SMM can undergo a twist to a anti-ferromagnetic SMM.  The twist can be effectively controlled
by two methods-(i) A simple substitution of one type of atom (in $Mn_{12}$ complexes a H atom)
 by another sterically demanding atom (again in $Mn_{12}$ complexes by a Me, Et, Ph, etc. atom) or by (ii) the other way this can be achieved
 is through the use of hydrostatic pressure\cite{brechin-press}. In this method external hydrostatic pressure is applied on the SMM leading to
 a twisting of the metallic core and thus a net change in magnetization. While the substitution method leads to a discrete change in the 
 magnetization the  pressure method leads to a continuous change. Therefore, the pressure method is much more amenable to the quantum  pumping
 mechanism wherein the pressure applied could be continuously changed albeit adiabatically leading to a adiabatic change in magnetization and 
 thus the exchange interaction $J$.

\section{Theory}
Graphene is a monatomic layer of graphite with a honeycomb lattice structure~\cite{graphene-rmp} that can be split into two
triangular sublattices $A$ and $B$. The electronic properties of graphene are effectively described by the massless Dirac equation. 
The presence of isolated Fermi points, $K$ and $K'$, in its spectrum, gives rise to two distinctive valleys.  We consider a sheet of
graphene on the $x$-$y$ plane. In Fig.~\ref{Fig1} we sketch our proposed system.
For a quantitative analysis we describe our system by the massless Dirac equation in presence of an embedded single magnetic molecule(SMM).
The Hamiltonian used to describe a SMM has the following terms:

\begin{equation}
H_{SMM}=-DS_{z}^{2} -J{\bf s}.{\bf S}
\end{equation}
The first term represents the Energy of the SMM. $D$ is an uniaxial anisotropy constant and $S_{z}$ is the z component of the 
spin of the Molecular magnet. The second term is most relevant to us since we deal with electron transport. The Dirac electrons 
in graphene interact with SMM only via the exchange term $-J{\bf s}.{\bf S}$. Further the magnitude in realistic SMM of $D$ is
very small as compared to $J$, $D=0.292 Kelvin=0.025$ meV while $J=100$ meV almost a thousand times larger\cite{barnas}. 
Though anisotropy is very important term in preserving the  intrinsic properties of SMM, it's effect on conduction electron
is hardly a small correction to energy, nothing else. The first term will only be relevant in electronic structure calculations
for electronic transport calculations it is not relevant and the only term of interest in the exchange coupling. 
Thus we only consider the second term in the subsequent analysis. 
We consider a single molecule magnet at $x=0$ and another electrostatic delta
potential at $x=a$ nm. The Hamiltonian of a Dirac  electron moving along x-direction can be written as
\begin{equation}
 H=\hbar v_F{\bf \sigma}.{\bf p}+J{\bf s}.{\bf S}\delta(x)+V\delta(x-a).
 \label{eq:H}
\end{equation}
The first term represent the kinetic energy term for graphene with $\bf \sigma$ the Paulli matrices  that operate on
the sub-lattices $A$ or $B$ and ${\bf p}=(p_x,p_y)$ the 2D momentum vector, second term is electron interacting with single molecule
magnet and third term is an electrostatic delta potential.  In the second term $J$ represents the exchange interaction 
which depends on the magnetization of the SMM and twisting the SMM changes the magnetization thus 
effectively changing $J$.  $ s$ represents the spin of the Dirac electron while $S$ represents spin of the SMM.
V represents strength of the adatom situated at a distance $a$ from SMM.
Here $\hbar, v_{F}$ (set equal to unity hence forth) are the Planck's constant and the energy
independent Fermi velocity for graphene. Eq.~\ref{eq:H} is valid near the valley $K$ in the Brillouin
zone and $\Psi=[\psi_{K}^{A}({\mathbf r}), \psi_{K}^{B}({\mathbf r})]$ is a spinor containing the electron fields in each
sublattice. The Hamiltonian for $K'$-valley can be obtained by replacing $p_y$ by $(-p_y)$ in Eq.~(\ref{eq:H}).
Because of this symmetry i.e; $H_{K}(p_y)=H_{K'}(-p_y)$, transport coefficients  will be same in both valleys.
So we are confining our discussion only in $K$-valley.

To calculate the quantum spin pumped currents we need to
introduce the basic theory of quantum pumping which is quite
well known as well as solve the scattering problem for electron
with spin (up or down) incident from either left or right. We
introduce them below:
 \subsection{Quantum pumped currents}
To calculate quantum pumped currents we proceed as follows:
Thus charge passing through lead $\mu$- to the left of Molecular magnet,  due to infinitesimal change of system parameters is given by-

\begin{eqnarray}
dQ_{\sigma \mu}(t)=e [\frac{dN_{\sigma \mu}}{dX_1} \delta X_{1}(t)+\frac{dN_{\sigma \mu}}{dX_2} \delta X_{2}(t)]
\end{eqnarray}

with the spin current transported in one period being-

\begin{eqnarray}
I_{\sigma \mu}=\frac{ew}{2\pi}\int_{0}^{\tau} dt [\frac{dN_{\sigma \mu}}{dX_1}\frac{dX_{1}}{dt}+
\frac{dN_{\sigma \mu}}{dX_2}\frac{dX_{2}}{dt}]\label{eq:Ipump}
\end{eqnarray}

In the above $\tau=2\pi/w$ is the cyclic period. The quantity
$dN_{\sigma\mu}/dX_i$ is the emissivity which is determined from
the elements of the scattering matrix, in the zero temperature limit
by -

\begin{eqnarray}
\frac{dN_{\sigma \mu}}{dX_i}=\frac{1}{2\pi}\sum_{\sigma\prime \nu} \Im (\frac{\partial s^{\sigma \sigma\prime}_{\mu \nu}}{\partial X_i}
s^{\sigma \sigma\prime *}_{ \mu \nu})
\label{eq:emis}
\end{eqnarray}

Here $s^{\sigma\sigma\prime}_{ \mu \nu}$ denote the elements of the scattering
matrix as denoted above, as evident $ \mu, \nu$ and $i$ can only
take values 1,2, while $\sigma, \sigma\prime$ takes values $\ua$ and $\da$ depending on
whether spin is up or down. The symbol ``$\Im$'' represents the
imaginary part of the complex quantity inside parenthesis.

The spin pump we consider is operated by adiabatically twisting the magnetic molecule thus modulating
the magnetic interaction between the magnetic molecule and scattered electrons ``J'' and
the strength of the ``delta'' function potential $V$, herein
$X_{1}=J=J_{0}+J_{p}\sin(wt)$ and
$X_{2}=V=V_{0}+V_{p}\sin(wt+\theta)$. A paragraph on the experimental
feasibility of the proposed device is given above the conclusion. As
the pumped current is directly proportional to $w$ (the pumping
frequency), we can set it to be equal to $1$ without any loss of
generality.

By using Stoke's theorem on a two dimensional plane, one can change
the line integral of Eq. (\ref{eq:Ipump}) into an area integral, see for details
Ref.\onlinecite{benjamin}-
\begin{eqnarray}
I_{\sigma \mu}=e\int_{A} dX_{1} dX_{2} [\frac{\partial}{\partial X_{1}}\frac{dN_{\sigma \mu}}{dX_2}-
\frac{\partial}{\partial X_{2}}\frac{dN_{\sigma \mu}}{dX_1}]
\label{stokes}
\end{eqnarray}

Substitution of Eq. (\ref{eq:emis}) into Eq. (\ref{stokes}) leads to,

\begin{eqnarray}
I_{\sigma \mu}=e\int_{A} dX_{1} dX_{2} \sum_{\sigma\prime=\ua,\da; \nu=1,2}
\Im (\frac{\partial s^{\sigma \sigma\prime*}_{\mu \nu}}{\partial X_1} \frac{\partial s^{\sigma \sigma\prime}_{\mu \nu}}{\partial X_2})
\end{eqnarray}
This current is for a particular angle of incidence ($\phi$), as the 
scattering amplitudes depend on $\phi$. So hereafter, we replace $I_{\sigma \mu}$ by $I_{\sigma \mu}(\phi)$.
If the amplitude of oscillation is small, i.e., for sufficiently weak
pumping ($\delta X_{i} \ll X_{i}$), we have,

\begin{eqnarray}
I_{\sigma \mu}(\phi)=\frac{ew\delta X_{1}\delta X_{2}sin(\theta)}{2\pi} \sum_{\nu=1,2} 
\Im (\frac{\partial s^{\sigma \sigma\prime*}_{ \mu \nu}}{\partial X_1} \frac{\partial s^{\sigma \sigma\prime}_{ \mu \nu}}{\partial X_2})
\end{eqnarray}
In the considered case of a magnetic molecule and delta function potential the case of very weak
pumping is defined by: $x_{p} \ll J_{0}=V_{0})$, and Eq.~8 becomes-

 \begin{eqnarray}
I_{\sigma \mu}(\phi)=I_{0} \sum_{\sigma\prime=\ua,\da; \nu=1,2}
\Im (\frac{\partial s^{\sigma\sigma\prime *}_{\mu \nu}}{\partial J} \frac{\partial s^{\sigma \sigma\prime *}_{\mu\nu}}{\partial V})
\end{eqnarray}
wherein,
\[I_{0}=\frac{ew x^{2}_{p} \sin(\theta)}{2\pi}\] .

As we consider only the pumped currents into lead 1(left of magnetic molecule), therefore
$\mu=1$. Further we drop the $\mu$ index in expressions below.

For weak pumping, we have the total pumped up-spin current given as follows:
\begin{eqnarray}
 I_{\ua}(\phi)&=&I_0\Big[{ \Im}\Big(\frac{\partial {s}^{\ua\ua *}_{11}}{\partial V}
 \frac{\partial s^{\ua\ua}_{11}}{\partial J}\Big)+{ \Im}\Big(\frac{\partial {s}^{\ua\da *}_{11}}{\partial V}
 \frac{\partial s^{\ua\da}_{11}}{\partial J}\Big)+{ \Im}\Big(\frac{\partial {s}^{\ua\ua *}_{12}}{\partial V}
 \frac{\partial s^{\ua\ua}_{12}}{\partial J}\Big)\nonumber\\&+&{ \Im}\Big(\frac{\partial {s}^{\ua\da *}_{12}}{\partial V}
 \frac{\partial s^{\ua\da}_{12}}{\partial J}\Big)\Big].
 \end{eqnarray}

 Similarly, we can calculate the spin down current for the case of weak pumping by replacing $\ua\rightarrow\da$ and vice-versa.
 
 For strong pumping, we have the total up-spin current given as:
 
  \begin{eqnarray}
I _{\ua}&=&\frac{ew}{2\pi}\int_{0}^{\tau} dt \big[\frac{dN_{\ua}}{dX_1}\frac{dX_{1}}{dt}+
\frac{dN_{\ua}}{dX_2}\frac{dX_{2}}{dt}], \mbox{with} \nonumber\\
\frac{dN_{\sigma \mu}}{dX_i}&=&\frac{1}{2\pi}[ \Im (\frac{\partial s^{\ua\ua}_{11}}{\partial X_i}
s^{\ua \ua *}_{11})+\Im (\frac{\partial s^{\ua\da}_{11}}{\partial X_i}s^{\ua \da *}_{11})+
\Im(\frac{\partial s^{\ua\ua}_{12}}{\partial X_i}s^{\ua \ua *}_{12})\nonumber\\&+&
\Im(\frac{\partial s^{\ua\da}_{12}}{\partial X_i}s^{\ua \da *}_{12})\big].
\end{eqnarray}
here we have dropped the lead index since we pump always to left lead (left to SMM)
and $s^{\sigma \sigma\prime *}_{\mu\nu}$ is complex conjugate of $s^{\sigma \sigma\prime}_{\mu\nu}$.
To obtain the total current, we integrate over $\phi$.
Then the total spin-up pumped current for both weak as well as strong pumping becomes: 
\begin{equation}\label{spin_up}
 I_{\ua}=\int_{-\pi/2}^{\pi/2}I_{\ua}(\phi)\cos(\phi)d\phi.
\end{equation}
 Similarly, we can calculate the spin down current for the case of weak pumping by replacing $\ua\rightarrow\da$ and vice-versa.
In the above equations the scattering amplitudes represent-
$s^{\ua\ua}_{11}=r_{\ua}$, reflection amplitude when spin-up electron is coming from the left side
and reflected to the spin-up state.\\
$s^{\ua\da}_{11}=r^{\prime}_{\ua}$, reflection amplitude when spin-down electron is coming from the left side
and reflected to the spin-up state\\
$s^{\ua\ua}_{12}=t_{\ua}$, transmission amplitude when spin-up electron is coming from the right side
and transmitted to the spin-up state.\\
$s^{\ua\da}_{12}=t^{\prime}_{\ua}$, transmission amplitude when spin-down electron is coming from the right side
and transmitted to the spin-up state.\\
Numerically, we have calculated $r_{\ua}$, $r^{\prime}_{\ua}$, $t_{\ua}$ and $t^{\prime}_{\ua}$ and substituted
in the above expression to obtain the spin-up pumping current.\\ 
{\em Effect of finite temperature:} So far our discussion is for zero temperature. The temperature effects could be easily captured by multiplying a
factor $[-df(E)/dE]$ with pumping current and integrating over energy of incident electron\cite{buttiker} i.e;
\begin{equation}\label{spin_up}
 I_{\sigma}=\int_{-\pi/2}^{\pi/2}\int_{0}^{\infty} \big[-\frac{df(E)}{dE}\big]I_{\sigma}(\phi)\cos(\phi)dE d\phi, \sigma=\ua,\da.
\end{equation}
where $f(E)$ is the Fermi-Dirac distribution function.
 
 \subsection{Solving the scattering problem}
 Let us consider an incident spin- up electron from left of magnetic molecule ($x<0$) with energy $E$.
The electron can be reflected or transmitted to spin-up/down state.
Then the spinors, for the angle of incidence $\phi$, in the various regions are given as:
\\
Region-I ($x<0$):\\
\begin{equation}
 \psi_{A}^{I}(x)=
 \left[\begin{array}[c]{c}
 (e^{ikx}+r_{\uparrow}e^{-ikx})\chi_m\\r_{\downarrow}e^{-ikx}\chi_{m+1}
 \end{array}\right]
\end{equation}
\begin{equation}
 \psi_{B}^{I}(x)=
\left[\begin{array}[c]{c}
 (e^{ikx+i\phi}-r_{\uparrow}e^{-ikx-i\phi})\chi_m\\-r_{\downarrow}e^{-ikx-i\phi}\chi_{m+1}
  \end{array}\right]
\end{equation}
 Region-II ($0<x<a$):\\
 \begin{equation}
 \psi_{A}^{II}(x)=
 \left[\begin{array}[c]{c}
 (a_{\uparrow}e^{ikx}+b_{\uparrow}e^{-ikx})\chi_m\\(a_{\downarrow}e^{ikx}+b_{\downarrow}e^{-ikx})\chi_{m+1}
  \end{array}\right]
\end{equation}
\begin{equation}
 \psi_{B}^{II}(x)=
 \left[\begin{array}[c]{c}
 (a_{\uparrow}e^{ikx+i\phi}-b_{\uparrow}e^{-ikx-i\phi})\chi_m
 \\(a_{\downarrow}e^{ikx+i\phi}-b_{\downarrow}e^{-ikx-i\phi})\chi_{m+1}
  \end{array}\right]
\end{equation}
and in Region-III ($x>a$):\\
\begin{equation}
 \psi_{A}^{III}(x)=
\left[\begin{array}[c]{c}
 t_{\uparrow}e^{ikx}\chi_m\\t_{\downarrow}e^{ikx}\chi_{m+1}
  \end{array}\right]
\end{equation}
\begin{equation}
 \psi_{B}^{III}(x)=
 \left[\begin{array}[c]{c}
 t_{\uparrow}e^{ikx+i\phi}\chi_m\\t_{\downarrow}e^{ikx+i\phi}\chi_{m+1}
  \end{array}\right].
\end{equation}

In the above equations, $\ua$ and $\da$ stand for spin-up and spin-down electron. Here,
 $r_{\ua}, r_{\da}$ and $t_{\ua}, t_{\da}$ are the reflection and transmission amplitudes respectively.
 Also, $k=E_{F}\cos(\phi)$ with $E_F(>0)$ being the Fermi energy. 
 $\chi$'s denote the eigen states of $S_z$ the z-component of spin operator for SMM, $ S_z\chi_m=m\chi_m$
with $m$ being the corresponding eigen value. The scattering is elastic
and the exchange interaction conserves the z-component of the total spin $S+s$.
The exchange operator in the Hamiltonian, ${\bf s.S}=s_zS_z+(1/2)(s^{-}S^{+}+s^{+}S^{-})$ acts
as a spin-flipper for electrons with different values of $s_z$ to those of $S_z$ while for same values it acts as a normal barrier.
$s^{-}S^{+}\left[\begin{array}[c]{c}
       1\\0
       \end{array}\right]\chi_m
=F\left[\begin{array}[c]{c}
         0\\1
        \end{array}\right]\chi_{m+1}$ and $s^{+}S^{-}\left[\begin{array}[c]{c}
       0\\1
       \end{array}\right]\chi_m
=F'\left[\begin{array}[c]{c}
         1\\0
        \end{array}\right]\chi_{m-1}$ with $F=\sqrt{(S-m)(S+m+1)}$ and $F'=\sqrt{(S+m)(S-m+1)} $.
Here, $s^{\pm}=s_{x}\pm s_{y}$ and $S^{\pm}=S_{x}\pm S_{y}$ are the raising and lowering operators.\\

In solving the scattering problem from a  delta potential the following two boundary conditions have to be met:
i) continuity of the wave functions at the boundary and ii) discontinuity of the energy at the boundary.
However, these boundary conditions work only if the system is described by the Schroedinger equation, but
does not work for Dirac material like graphene. For Dirac equation, the wave functions on either side of delta potential are not
continuous across the boundary. 

At x=0, while taking integration on the  both sides of the Dirac equations, $H\psi=E\psi$, one is stuck at the following
integration:
\begin{equation}
 \bar{\psi}(0)=\int_{x=0^-}^{x=0^+}\psi\delta(x)dx
\end{equation}
because of the discontinuity of wave functions at the boundary. However, there
is a standard way to avoid this difficulty, which has been widely used by many authors,
by taking the average, i.e., 
\begin{equation}
 \bar{\psi}(0)=\frac{1}{2}[\psi (x=0^+)+\psi(x=0^-)],
\end{equation}
where the delta function potential mentioned  is
not an exact delta function potential but a point like interaction\cite{griffiths,bose}.
The above idea can be deployed at the boundary x=0, which leads to the two equations as
\begin{eqnarray}\label{ba}
 -i\hbar v_{F}[\psi_{B}^{II}(x=0^+)&-&\psi_{B}^{I}(x=0^-)]\nonumber\\&=&\frac{J}{2}\vec{s}.\vec{S}
 [\psi_{A}^{II}(x=0^+)+\psi_{A}^{I}(x=0^-)].
\end{eqnarray}
\begin{eqnarray}\label{ab}
 -i\hbar v_{F}[\psi_{A}^{II}(x=0^+)&-&\psi_{A}^{I}(x=0^-)]\nonumber\\&=&\frac{J}{2}\vec{s}.\vec{S}
 [\psi_{B}^{II}(x=0^+)+\psi_{B}^{I}(x=0^-)].
\end{eqnarray}
\begin{figure*}
  \centering
 \subfigure[Quantum pumping current Vs. Energy of incident electron for weak pumping]{ \includegraphics[width=0.49\textwidth]{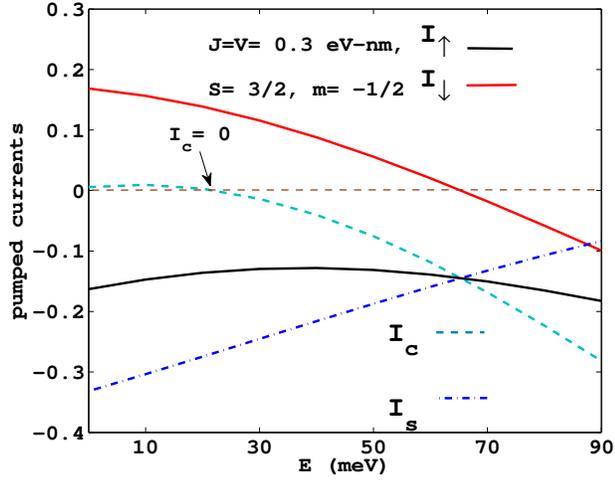}}
   \subfigure[Quantum pumping current Vs. strength of the delta like point interaction for weak pumping]{\includegraphics[width=.49\textwidth]{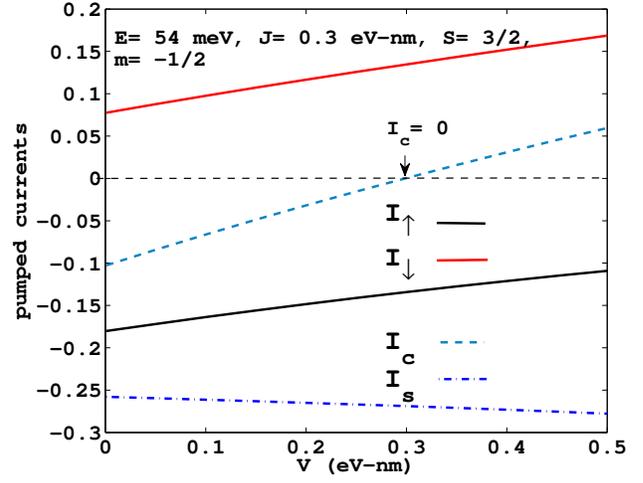}}
    \centering
 \subfigure[Weak Pumping: Quantum pumping current Vs. Strength of the molecular magnet]{  \includegraphics[width=.49\textwidth]{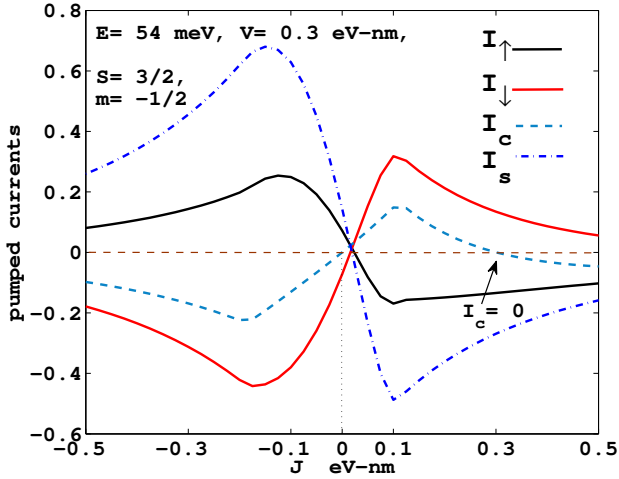}}
\subfigure[Strong Pumping: Quantum pumping current Vs. Strength of the molecular magnet]{   \includegraphics[width=.49\textwidth]{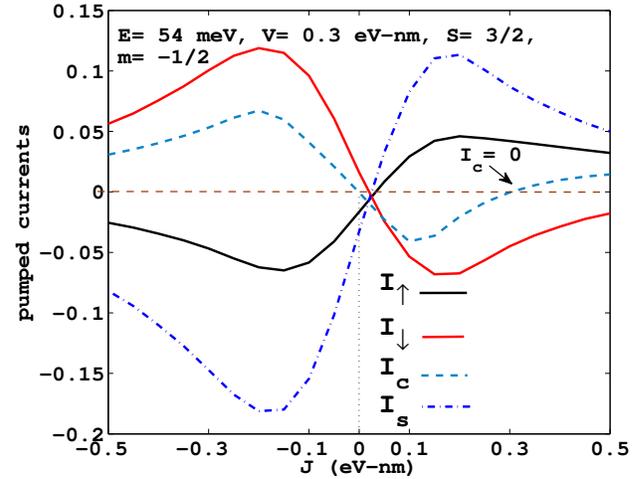}}
\caption{Pumped spin currents at zero temperature, other parameters are mentioned in Figure.}
\end{figure*}
After substituting the wave functions in Eq.~(\ref{ba}) and Eq.~(\ref{ab}),
\begin{eqnarray}\label{first}
  a_{\ua}(i\alpha m &+& e^{i\phi})+a_{\da}(i\alpha F )+b_{\ua}(i\alpha m-e^{-i\phi})+
 b_{\da}(i\alpha F )\nonumber\\&+&r_{\ua}(i\alpha m+e^{-i\phi})+r_{\da}(i\alpha F )
 =e^{i\phi}-i\alpha m,
\end{eqnarray}
\begin{eqnarray}
  a_{\ua}(i\alpha F)&+&a_{\da}[e^{i\phi}-i\alpha (m+1)]+b_{\ua}(i\alpha F)
 -b_{\da}[i\alpha (m+1)+e^{-i\phi}]\nonumber\\&+&r_{\ua}(i\alpha F)+r_{\da}[e^{-i\phi}-i\alpha (m+1)]
 =-i\alpha F,
\end{eqnarray}
\begin{eqnarray}
 a_{\ua}(1&+&i\alpha me^{i\phi})+a_{\da}(i\alpha F e^{i\phi})+b_{\ua}(1-i\alpha me^{-i\phi})-
 b_{\da}(i\alpha F e^{-i\phi})\nonumber\\&-&r_{\ua}(1+i\alpha me^{-i\phi})-r_{\da}(i\alpha F e^{-i\phi})
 =1-i\alpha m e^{i\phi},
\end{eqnarray}
and
\begin{eqnarray}
 &&a_{\ua}(i\alpha Fe^{i\phi})+a_{\da}[1-i\alpha (m+1) e^{i\phi}]\nonumber\\&-&b_{\ua}(i\alpha Fe^{-i\phi})+
 b_{\da}[1+i\alpha (m+1) e^{-i\phi}]\nonumber\\&-&r_{\ua}(i\alpha Fe^{-i\phi})-r_{\da}[1-i\alpha (m+1) e^{-i\phi}]=i\alpha F e^{i\phi}.
\end{eqnarray}
Here, $\alpha=J/(2\hbar v_F)$. The mixing of the spin-up and spin-down components in the above equations are 
attributed to the exchange operator ${\mathbf s}\cdot{\mathbf S}$.\\
At x=a, the boundary conditions are given as:
\begin{eqnarray}\label{ba1}
 -i\hbar v_{F}[\psi_{B}^{III}(x=a^+)&-&\psi_{B}^{II}(x=a^-)]\nonumber\\&=&\frac{V}{2} [\psi_{A}^{III}(x=a^+)+\psi_{A}^{II}(x=a^-)]
 \nonumber
\end{eqnarray}
and
\begin{eqnarray}\label{ab1}
 -i\hbar v_{F}[\psi_{A}^{III}(x=a^+)&-&\psi_{A}^{II}(x=a^-)]\nonumber\\&=&\frac{V}{2} [\psi_{B}^{III}(x=a^+)+\psi_{B}^{II}(x=a^-)],
 \nonumber
\end{eqnarray}
lead to
\begin{eqnarray}
 a_{\ua}e^{ika}(iG-e^{i\phi})+b_{\ua}e^{-ika}(iG+e^{-i\phi})+t_{\ua}e^{ika}(e^{i\phi}+iG)=0,\nonumber\\
\end{eqnarray}
\begin{equation}
 a_{\da}e^{ika}(iG-e^{i\phi})+b_{\da}e^{-ika}(iG+e^{-i\phi})+t_{\da}e^{ika}(e^{i\phi}+iG)=0,\\
\end{equation}
\begin{equation}
 a_{\ua}e^{ika}(-1+iGe^{i\phi})-b_{\ua}e^{-ika}(1+iGe^{-i\phi})+t_{\ua}e^{ika}(1+iGe^{i\phi})=0,
\end{equation}
and
\begin{equation}\label{last}
 a_{\da}e^{ika}(-1+iGe^{i\phi})-b_{\da}e^{-ika}(1+iGe^{-i\phi})+t_{\da}e^{ika}(1+iGe^{i\phi})=0,
\end{equation}
where, $G=V/(2\hbar v_F)$.\\
The eqns. (\ref{first}-\ref{last}) contain 8 unknown probability amplitudes, which
can be solved numerically to confirm $\mid t_{\ua}\mid^2+\mid r_{\ua}\mid^2+\mid t_{\da}\mid^2+\mid r_{\da}\mid^2=1$.
Similarly for the case of spin-down incident electron from the left side, we can get scattering amplitudes.
This procedure can be repeated appropriately for spin-up/down electron coming from right side.
 
 \section{Results and discussions}
 Different components of quantum pumped currents i.e; spin-up ($I_\ua$), spin-down ($I_\da$) and net spin current
 ($I_s=I_\ua-I_\da$) and charge current $(I_c=I_\ua+I_\da$) are obtained numerically using Eq.~12 at zero temperature as shown in figures 2 (a)-(d).

Figure 2(a) and 2(b) show the quantum pumped currents $(I_\ua,I_\da,I_s,I_c)$ versus energy of the incident electron $(E)$ and 
 strength of adatom/line defect for weak pumping, without disorder. The following numerical parameters are used (mentioned in the figure also):  the spatial separation between  SMM and the adatom $a= 10$ nm,
 the strength of exchange interaction between electron and  the molecular magnet $J=0.3$ eV-nm, $m = -1/2$ and spin of the molecular magnet $S=3/2$. The strength of  line defect potential $V=0.3$ eV-nm and  the strength of the time varying modulations on top of V and J are taken as $x_p=0.05$ eV-nm in strong pumping case. From figure 2(a), it is seen that for a suitable
 energy $E$ ($=54$ meV), the total charge current completely disappears leaving behind a pure spin-current.
 Fig. 2(b) shows that individual spin pumped currents slowly varying with the increase of ordinary potential.
 The $I_{\ua}$ is decreasing while $I_{\da}$ is increasing with V. The pumped currents in weak pumping regime
are in units of $I_{0}$ as in Eq. 8, while in strong pumping regime are in units of $ew/2\pi$. 	Another important point in
the weak pumping case is that, by suitably choosing E, $I{\da}$ can
be completely suppressed to pump only $I\ua$ a spin up selective
current.

 Figures 2(c) and 2(d) are plotted to check how the strength of the exchange interaction (J) affects
 the different pumped currents. We have  taken  energy of incident electron ($E=54$ meV) for which pure spin current was observed.  From figure 2(c), we see that the pumped currents attain a maximum around $J= 0.1$ and $-0.13 $ eV-nm and then start diminishing towards zero, and pure spin current appears at  $J= 0.3$ eV-nm, and 
 spin selective current $I_{\da}$ appears for  $J>0.5$ eV-nm.
 However, in strong pumping as shown in figure (d), pure spin current appears at the same position at $J=0.3$ eV-nm.
 For both weak and strong pumping plots, we have chosen the phase difference between two modulations
as $\pi/2$. One important fact which is clearly noticeable is the fact that `J' acts as a current switch. By changing J from positive to negative, all the pumped currents change sign. This shows that the magnetization of  SMM controls the direction of current flow in graphene. Since, we aim to control this by the twist, effectively twisting the SMM changes  the direction of spin currents. This is a key result of our work.
\subsection{Effect of Disorder and temperature}
In this subsection, we discuss the effects of disorder distributed randomly in the system. We have modeled the present
device in such way that SMM and adatom/line defect are at the extreme ends of the sample and disorder is confined between SMM and adatom/line defect.
Here each disorder is considered to be a delta potential (point interaction as mentioned before). 
We solve this problem by using transfer matrix approach. The presence of multiple delta potentials 
causes a number of confined regions between  SMM and the adatom. The general form
of the wave function in each region can be written as:

\begin{figure}
\begin{center}
\includegraphics[width=.5\textwidth,height=40mm]{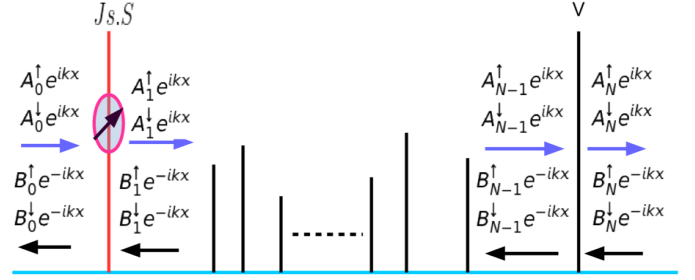}
\caption{A typical diagram of the disorder added device in which SMM and adatom is separated by a distance $a$ nm.}
\end{center}
\end{figure}

\begin{figure*}
 \subfigure[Quantum pumped currents in the weak pumping regime vs. spin of the SMM for disordered graphene.]{ \includegraphics[width=.49\textwidth]{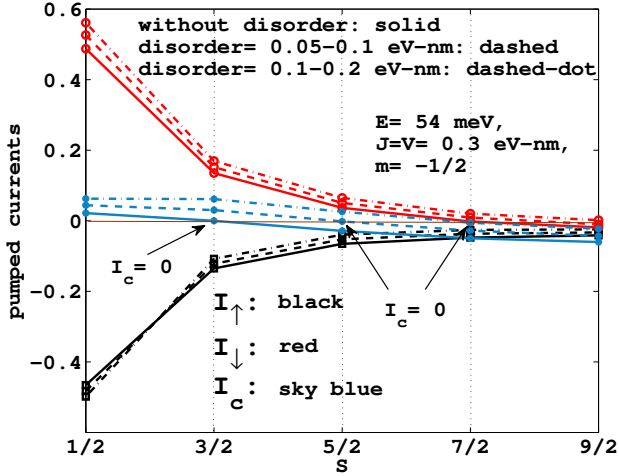}}
 \subfigure[Quantum pumped currents in the weak pumping regime vs. J for disordered graphene.]{ \includegraphics[width=.49\textwidth]{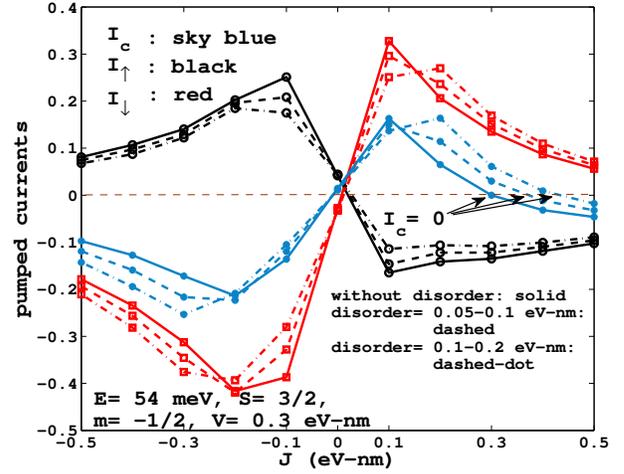}}
\caption{Pumped spin currents in presence of disorder.}
\end{figure*}

\begin{figure}
\begin{center}
{ \includegraphics[width=.49\textwidth]{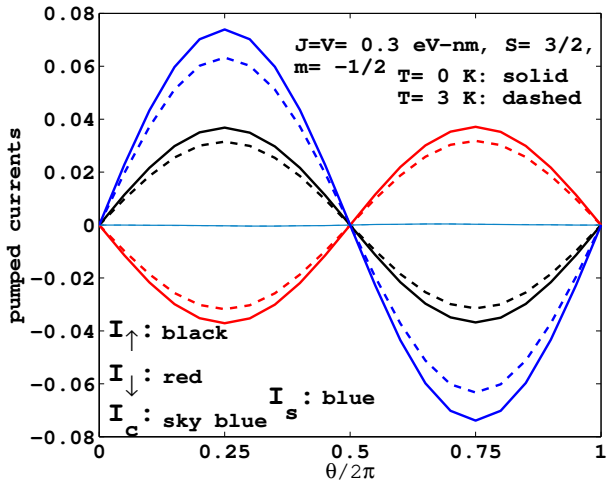}}
\caption{Quantum pumping current Vs. phase angle between two modulations for strong pumping. Solid line is for zero temperature case while dashed line is for finite temperature.}
\end{center}
\end{figure}
\begin{equation}
 \psi_{n}(x)=
 \left[\begin{array}[c]{c}
 (A^{\uparrow}_{n}e^{ikx}+B_{n}^{\uparrow}e^{-ikx})\chi_m\\(A^{\downarrow}_{n}e^{ikx}+B^{\downarrow}_{n}e^{-ikx})\chi_{m+1}
 \end{array}\right].
\end{equation}
Here $n=0,1,2,3...(N-1),N$ corresponding to different regions bounded by the delta potentials, as shown in Fig.~3.
The above wave function is for A-sublattice, the phase factor $\exp(\pm i\phi)$ is multiplied with
transmission (reflection) amplitude to get the same for B-sublattice.
The next step is to find out the  total transfer matrix which connects the wave function amplitudes between extreme left and right.
To do so, first we find  transfer matrix across SMM i.e; between region $`n=0'$ and $`n=1'$ as in Fig. ~3,
\begin{equation}\label{T-matrix}
 \left[\begin{array}[c]{c}A^{\uparrow}_{1}\\A^{\downarrow}_{1}\\B^{\uparrow}_{1}\\B^{\downarrow}_{1}\end{array}\right]=
M^{[1,0]}\left[\begin{array}[c]{c}A^{\uparrow}_{0}\\A^{\downarrow}_{0}\\B^{\uparrow}_{0}\\B^{\downarrow}_{0}\end{array}\right],
\end{equation}
where $M^{[1,0]}$ is the transfer matrix across SMM, expressed as $M^{[1,0]}=M_{1}/M_{0}$ with
\begin{equation}
 M_{0}=\left[\begin{array}[c]{cccc} \xi- igm& -igF &igm-\xi_c & -igF \\-igF &\xi+ig(m+1)& -igF &-ig(m+1)-\xi_c\\
            1-igm \xi & -igF\xi &1+igm \xi_c &igF\xi_c\\-igF \xi& 1+ig(m+1)\xi& igF\xi_c& 1-ig(m+1)\xi_c
           \end{array}\right]
\end{equation}
and
\begin{equation}
 M_{1}=\left[\begin{array}[c]{cccc} \xi+ igm& igF &igm-\xi_c & igF \\igF &\xi-ig(m+1)& igF &-ig(m+1)-\xi_c\\
            1+igm \xi & igF\xi &1-igm \xi_c &-igF\xi_c\\igF \xi& 1-ig(m+1)\xi& -igF\xi_c& 1+ig(m+1)\xi_c
           \end{array}\right]
\end{equation}
with $\xi=\exp(i\phi)$ and $\xi_c=\exp(-i\phi)$.
Similarly we can get the transfer matrix across any ordinary potential, for example, the  transfer matrix between $`n=N'$ and $`n=N-1'$ as
\begin{equation}
 \left[\begin{array}[c]{c}A^{\uparrow}_{N}\\A^{\downarrow}_{N}\\B^{\uparrow}_{N}\\B^{\da}_{N}\end{array}\right]=
M^{[N,N-1]}\left[\begin{array}[c]{c}A^{\uparrow}_{N-1}\\A^{\da}_{N-1}\\B^{\uparrow}_{N-1}\\B^{\da}_{N-1}\end{array}\right],
\end{equation}
where $M^{[N,N-1]}$ is the transfer matrix across adatom, expressed as $M^{[N,N-1]}=M_{N}/M_{N-1}$ with
\begin{equation}
 M_{N-1}=\left[\begin{array}[c]{cccc}\xi- iG& 0 &iG-\xi_c & 0 \\0 &\xi-iG& 0 &iG-\xi_c\\
            1-iG \xi & 0&1+iG \xi_c &0\\0& 1-iG\xi& 0& 1+iG\xi_c
           \end{array}\right]
\end{equation}
and
\begin{equation}
 M_{N}=\left[\begin{array}[c]{cccc} \xi+ iG& 0 &iG-\xi_c & 0 \\0 &\xi+iG& 0 &iG-\xi_c\\
            1+iG \xi & 0 &1-iG \xi_c &0\\0& 1+iG\xi& 0& 1-iG\xi_c
           \end{array}\right].
\end{equation}
Since the adatom is modeled as a delta function potential and disorder is modeled too as randomly distributed 
sequence of delta potentials with random strengths  the transfer matrix for any arbitrary  interface   between adatom 
and SMM has also the same matrix elements as $M^{[N,N-1]}$.

After some straight forward algebra, the connection between the wave function amplitudes of extreme left and right is found to be\cite{griffiths}
\begin{equation}
 \left[\begin{array}[c]{c}A^{\uparrow}_{N}\\A^{\da}_{N}\\B^{\uparrow}_{N}\\B^{\da}_{N}\end{array}\right]=
M\left[\begin{array}[c]{c}A^{\uparrow}_{0}\\A^{\da}_{0}\\B^{\uparrow}_{0}\\B^{\da}_{0}\end{array}\right],
\end{equation}
where
\begin{equation}
       M=M^{[N,N-1]}M_{free}^{[N-1]}M^{[N-1,N-2]}M_{free}^{[N-2]}.....M_{free}^{[1]}M^{[1,0]}
      \end{equation}
with $M^{n}_{free}$, the propagation matrix between any two successive delta potential, is given by
\begin{equation}
 M^{n}_{free}=\left[\begin{array}[c]{cccc}  e^{ikd_n}& 0& 0 &0 \\0 &e^{ikd_n} & 0 &0\\
            0 & 0 &e^{-ikd_n} &0\\0& 0& 0&e^{-ikd_n}
           \end{array}\right].
\end{equation}
Here, $d_n$ is the separation between two consecutive delta potentials.
To calculate the reflection and transmission amplitudes, we shall use the scattering matrix (S-matrix), 
which is connected to transfer matrix as\cite{griffiths}
\begin{eqnarray}
 S&=&\frac{1}{(M_{22})_{2\times 2}}\left[\begin{array}[c]{cc}  (M_{21})_{2\times 2}& 1 \\ \det(M) &(M_{12})_{2\times 2}
                       \end{array}\right],\nonumber\\
                       M&=&\left[\begin{array}[c]{cc}  M_{11}& M_{12} \\ M_{21} & M_{22}
                       \end{array}\right]=\left[\begin{array}[c]{cccc}  m_{11}& m_{12}&m_{13}&m_{14}\\ m_{21}& m_{22}&m_{23}&m_{24} \\ m_{31}& m_{32}&m_{33}&m_{34} \\ m_{41}& m_{42}&m_{43}&m_{44}
                       \end{array}\right].
\end{eqnarray}
The reflection amplitude (to the left, as we are calculating pumped currents in the left lead)
\begin{equation}
 r_{_l}=-\frac{M_{21}}{M_{22}}=\left[\begin{array}[c]{cc}  r_{\uparrow\uparrow}& r_{\uparrow\downarrow} \\ 
 r_{\downarrow\uparrow} &r _{\downarrow\downarrow}
                       \end{array}\right]
\end{equation}
and transmission amplitude from right to left is
\begin{equation}
 t_{r}=\frac{1}{M_{22}}=\left[\begin{array}[c]{cc}  t_{\uparrow\uparrow}& t_{\uparrow\downarrow} \\ 
 t_{\downarrow\uparrow} &t _{\downarrow\downarrow}
                       \end{array}\right].
\end{equation}
The $t_r$ and $r_{_l}$ can be directly used in Eq. (\ref{spin_up}) to obtain the  quantum pumped currents. We must
mention here that disorder free pumping current can also be recovered from here by using transfer matrix as $M=M^{[2,1]}M_{free}M^{[1,0]}$,
where $M^{[2,1]}$ and $M^{[1,0]}$ would become the transfer matrix across adatom and SMM, respectively.

 Disorder effect is  shown in Fig.~4(a).  Herein we plot the pumped currents as function of the different 
 spin states of SMM in the weak pumping regime. There is pure spin current for $S = 3/2$, in
the figure $m=-1/2$. To include disorder, we have chosen the random spacing between any two 
potentials  in the range $1-1.5$ nm. The strength of the  potential is also random and ranges from $50-100$ meV-nm (dashed line) to  $100-200$ meV-nm (dashed-dot line).
We have considered the pumped currents averaged  over 1000 realizations.   One can see that disorder has a limited effect effect on the pure spin current.
The position of pure spin current is shifted from $S=3/2$ (without disorder ) to $S=5/2$ (disorder:$50-100$ meV-nm) and finally  $S=7/2$ (disorder:$100-200$ meV-nm) however pure spin currents aren't killed off.  In Fig. 4b we see the effect of disorder on pumped currents plotted as function of the magnetization of SMM. We again see position of occurrence of pure spin current ($I_{c}=0$) changes from $J=0.3$  to $0.45$ eV-nm as one increases disorder.  However, disorder has no effect on magnetization switching of pumped spin currents showing the resilience of the magnetization switching to disorder.

Finally, in figure 5 we plot the pumped currents as function of the phase difference between modulated parameters. We see that pumped current attains 
maximum at $\theta=\pi/2$ and minima around
$\theta=0,\pi$. The pure spin current is maintained throughout the whole range of $\theta$.
The temperature effect is shown in  the same figure, which shows a small damping in amplitudes of
the individual spin currents. One can see that temperature has no noticeable effect on the pure spin currents apart from a
slight diminishing of the magnitude. To conclude this section, pure spin currents in graphene are immune to any temperature increase apart from decrease in magnitude while disorder has a small effect as it shift the parameter regime for occurrence of pure spin currents although it cannot kill it off.

  \section{Pumping Vs. Rectification}
  A major issue which was flagged right from the early days of quantum pumping was whether the Switkes experiment\cite{switkes}
  was a real demonstration of quantum pumping, since the pumped current was observed to be symmetric with respect to Magnetic 
  field reversal just like the two terminal conductance\cite{brou-rect}. However since pumped currents are functions of 
  scattering amplitudes and not scattering probabilities they should have no particular symmetry with respect to magnetic 
  field reversal unless the system itself had some particular symmetry\cite{shut}. As the Switkes expt. system did not
  possess any particular symmetry it was quickly recognized that the current attributed as a a pumping current was in effect
  a rectified current which depend on the Conductance of the system\cite{benj}. However there could have been a pumped current
  which was masked by  the rectified currents. The origin of rectified currents is because experimentally at the nanoscale it
  is difficult to control time varying parameters. Most naturally time varying parameters couple to input and output leads and 
  instead of only pumping a current there is in addition a transport current defined the net conductance through the system. 
  So any quantum pumping at the nanoscale will have rectified currents and therefore it become imperative to have a scheme to differentiate these currents.  The rectified spin up current is defined as:
  $I_{rect} = \frac{w}{2\pi} R \int_{S} dX_{1}dX_{2} (C_{1}
  \frac{\partial G_{\ua}}{\partial X_{1}}-C_{2}\frac{\partial G_{\ua}}{\partial X_{2}})$, $X_{i}$'s are the modulated parameters. 
  In the weak pumping regime  we have- $I_{rect} =I_{0} (C_{1} \frac{\partial G_{\ua}}{\partial X_{1}}-C_{2}\frac{\partial G_{\ua}}{\partial X_{2}})$,
  with $I_{0}=we^{2}\sin(\theta)\delta X_{1}\delta X_{2}R/4\pi^{2}\hbar$.
  
\begin{figure}
    \includegraphics[width=.5\textwidth,height=20em]{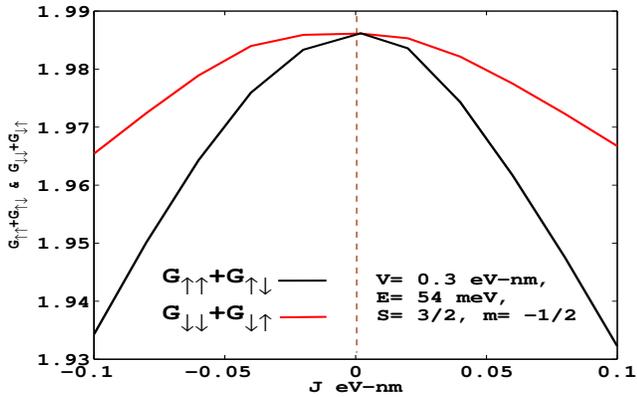}
    \caption{Conductance (spin up and spin down) vs. J. For small J values G's are symmetric. Compare with Fig 2(c) and (d).
    Pumped currents up and down spin are asymmetric.  }\label{rect}
  \end{figure}

  In Fig. \ref{rect} We can see clearly that the Conductance (both spin-up as well as spin-down)
  are symmetric with respect to small values of J. Thus unlike pumped currents whose direction can be changed by changing the magnetization from positive to negative, the conductance on the other hand shows no such effect. The pumped currents are completely 
  asymmetric as function of J as seen in Fig. 2(c) and (d). Thus even if rectified currents will be present in the system the pumped
  spin current will be distinguished because of their asymmetric nature with respect to magnetization reversal.

\section{Experimental realization and Conclusions}

The experimental realization of our pure spin current pumping device shouldn't be too difficult.
As already outlined in the last paragraph of the introduction of this paper adiabatically
modulating the pressure applied on the single molecule magnet would entail a corresponding 
adiabatically modulated magnetization of SMM. The second 
adiabatically modulated parameter of the device is an adatom placed ``a'' distance apart from SMM. 
The adatom is modeled as a delta function like point interaction similarly embedded in graphene.  
A gate voltage applied to the adatom can change the potential felt by electrons scattered from it. 
When the gate voltage itself is adiabatically modulated in time we have all the ingredients for the
quantum pumping of pure spin currents and spin selective currents. 
Alternatively, an extended line defect can be created instead of adatom, which can be 
controlled experimentally\cite{defect1,defect2,defect3}. Moreover, one can also use  a thin potential barrier experimentally which can be theoretically modeled as a delta like potential. Similarly, the single molecule magnet is infact a large molecule with host of atoms ranging from 30-100 atoms, these atoms are arranged not only having vertical but also horizontal extension i.e., a  single molecule magnet will have  sufficient extension along transverse direction.  Mention may be made of Ref.\onlinecite{bogani} on single molecule magnets which exemplifies the situation envisaged.

To conclude we have shown that notwithstanding the fact that spin transport via spin orbit effect is 
almost impossible to be observed in graphene, we have in a novel manner pumped pure spin currents and spin 
selective currents in graphene via embedding it with a single molecule magnet. The study of pure spin currents
in graphene via embedded SMM will be extended to spin correlations and whether one can generate entangled spin
currents which will have potential impact on quantum information processing\cite{benj-smm} in a subsequent work.

\end{document}